\documentclass[conference]{IEEEtran}
\usepackage{times}
\usepackage[T1]{fontenc}
\usepackage[greek,english]{babel}
\usepackage[cmex10]{amsmath}
\usepackage{graphicx}
\usepackage{enumerate}
\usepackage{amsmath,amssymb}
\usepackage{adjustbox}
\usepackage{enumitem}
\usepackage{color}
\usepackage{float}
\usepackage[noadjust]{cite}
\usepackage[justification=centering]{caption}
\usepackage{epstopdf}

\usepackage{url}
\begin{document}

\title{Decoding and File Transfer Delay Balancing in Network Coding Broadcast} 

\author{
    \IEEEauthorblockN{Emmanouil Skevakis}
    \IEEEauthorblockA{Department of Systems and Computer Engineering\\ Carleton University, \\
Ottawa, Ontario
K1S 5B6 Canada
    \\eskevakis@sce.carleton.ca}
    \and
     \IEEEauthorblockN{Ioannis Lambadaris}
    \IEEEauthorblockA{Department of Systems and Computer Engineering\\ Carleton University, \\
Ottawa, Ontario
K1S 5B6 Canada
    \\ioannis@sce.carleton.ca}
}

\maketitle

\begin{abstract}

Network Coding is a packet encoding technique which has recently been shown to improve network performance (by reducing delays and increasing throughput) in broadcast and multicast communications. The cost for such an improvement comes in the form of increased decoding complexity (and thus delay) at the receivers end. Before delivering the file to higher layers, the receiver should first decode those packets.% This is particularly limiting in applications where part (or parts) of the transmitted file can be useful to the higher layers. A parameter that affects both the completion time of a file and the decoding delay at the receivers end, is the Coding Window size $K$ (i.e. the number of packets that will produce an encoded packet).

%In this work, we model and analyse a system where a base station broadcasts a single file to $N$ receivers, when employing Random Linear Network Coding (RLNC). RLNC is a simple form of network coding and is asymptotically optimal in terms of broadcast completion time (\cite{1}). A heuristic is proposed and mathematically modelled, in order to reduce the completion time of the file, when coding is not performed over the whole file. Moreover, in the case of traffic with delay constraints% (as a percentage of the minimum completion time under RLNC), we obtain the minimum Coding Window size that satisfies those constraints. This way, the decoding delay at the receivers end is reduced, while guaranteeing a maximum completion time.

In our work we consider the broadcast transmission of a large file to $N$ wireless users. The file is segmented into a number of blocks (each containing $K$ packets - the Coding Window Size). The packets of each block are encoded using Random Linear Network Coding (RLNC). We obtain the minimum coding window size so that the completion time of the file transmission is upper bounded by a used defined delay constraint.
 
\end{abstract}

\section{Introduction}
Over the past decades, with the constantly increasing use of cellular and wireless networks for data transmission, the efficient utilization of the network and its resources has become crucial. Bandwidth-intensive applications such as video and music downloading and delay sensitive applications such as real-time video streaming and IPTV are widely deployed in wireless, unreliable, networks. Most of these applications involve the need of transmission of packets from one (or more) senders to multiple receivers. This has intensified the need for more reliable data transmissions with increased data rates in such scenarios.

Network Coding may improve the network performance in such cases and contribute towards achieving these goals (%\cite{3},
\cite{2},\cite{7},\cite{8}). In the standard approach for data transmission, information is transmitted to a receiver based on a scheduling policy and the channel conditions. With network coding, information can be transmitted to multiple receivers simultaneously even when each of the receivers expect different packets. This can be greatly beneficial in multicast and broadcast transmissions, where the same information must be sent to a large number of receivers. In network coding, each encoded packet is generated from a given number $K$ of packets (also know as the coding window) based on an encoding scheme. Most of the schemes introduce redundancy to the network, thus increasing reliability, without decreasing the network performance.

Many forms of network coding can be found in the literature, each one with its own benefits and drawbacks. Two of the most promising techniques are Random Linear Network Coding (RLNC) and Instantly Decodable Network Coding (IDNC). RLNC is one of the simplest forms of network coding that can approach system capacity with negligible feedback overhead \cite{4}. RLNC linearly combines packets within the coding window. After the successful reception of $K$ such packets (given that the linear combinations they represent are independent) the receivers can successfully decode the packets, with Gaussian elimination. This technique achieves asymptotically optimal completion time of a block of packets and higher throughput than any scheduling strategy (\cite{7},\cite{1}). However, the receivers must have received $K$ packets before the decoding process thus the decoding delay increases. Moreover, the authors in \cite{5} have proven that the coding window size $K$ has to scale with the number of receivers resulting in increased decoding complexity and decoding delay in large networks. IDNC is another form of network coding for minimizing decoding delay (\cite{12}). Different concepts of IDNC can be found in the literature depending on the application requirements for which they were developed (\cite{10},\cite{11}). The main advantage of IDNC over RLNC is the reduction of decoding delay (instant decoding by some or all receivers) at the cost of increased block completion time \cite{13}.

Relevant research by Eryilmaz focused on quantifying the gains of network coding in terms of throughput and delay (\cite{1},\cite{5}) in an unreliable (e.g., wireless) single-hop broadcast network. The improvement in network performance of network coding versus traditional scheduling strategies is proven and both asymptotic and close form expressions for the delay and throughput are provided. The effects of delay constrained traffic on the user admission rate are analysed in \cite{15} as an extension of the previous work. In \cite{7}, the authors focus on how must the Coding Window size $K$ scale when the number of receivers is increased. In such a case, the distribution of the delay is characterized. 

In \cite{16}, two schemes were proposed and analysed in order to minimize the decoding delay and feedback overhead while keeping the throughput intact. In \cite{17}, the randomness is dropped from the encoding procedure and with the integration of an ARQ strategy, the successful decoding is guaranteed. The authors of \cite{14} combine RLNC and IDNC through partitioning, in order to improve throughput, decoding delay, coding complexity and feedback frequency. Koller et al \cite{18}, investigates the optimum number of blocks that the source must use, given the file size and a finite Galois field size from which the coefficients are picked, when using RLNC. In \cite{19}, the authors try to find a linear network coding scheme that outperforms RLNC in terms of minimum average packet decoding delay while keeping the throughput intact. Their work is very interesting but it assumes already received packets by the receivers from previous transmissions and erasure-free NC transmissions.

The system we focus on in this paper is similar with the one studied in \cite{1}. The main difference from \cite{1} and \cite{5} lies in the Coding Window Size $K$. In our work, we are interested in a) \textit{finding the mean file transfer completion time (broadcast to $N$ receivers) when RLNC is not performed over the whole file}  and b) \textit{find a relationship between the coding window size ($K$) and the expected file transfer completion time}. In particular, we will develop a closed form formula for the minimum Coding Window Size $K$ in the case of a user defined delay constraint. The constraint will be in the form of the relative increase in the delay over the smallest achievable (optimal) delay. This objective might seem similar with the one in \cite{18}, but our system has some fundamental differences. First, in \cite{18}, the authors assume that the receivers will, on average, need more than $K$ packets for successful decoding. In our work, we assume a large enough field size (from where the coefficients for the linear combinations are picked) thus guaranteeing linear independence. As a result, only $K$ encoded packets are needed for successful decoding. Second, in \cite{18}, the file is split into blocks and coding is performed \textit{over those blocks} (i.e. the encoded block is the sum of each block multiplied by a coefficient). In our work, coding is performed \textit{within each block}. By doing so, packets of earlier blocks are made available (for delivery to the higher protocol layers) faster. This way, we keep the advantages of asymptotically optimal completion time of RLNC while reducing the decoding delay and complexity at the receiver.

The remainder of the paper is organized as follows : In Section II, the system model is introduced. In Section III, mathematical formulas for the completion time of a file for a given coding window size of $K$ are derived. In Section IV, we develop the formula for the smallest Coding Window Size $K$ under user defined delay constraints. In Section V we show our results and comparisons. Finally, in section VI we provide our conclusions and future research directions.

\section{System Model}

Our system is a one hop transmission system, where a base station transmits a single file (containing $F$ packets) to $N$ receivers. The connection between the base station and each receiver is described by a randomly varying ON/OFF channel, where the $i^{th}$ receiver's channel state is represented by a Bernoulli random variable with mean $p_i$. In our simulations and mathematical formulations (section III and IV), we assume, for simplicity, that $p_i$ = $p$, $\forall i \in \{1, ... , N\}$. However, extensions to non-symmetric channels can be easily implemented. The Bernoulli r.v's are independent across time and across receivers. Furthermore, we assume that a base station, at each time slot, has knowledge of the connected receivers and in this case, a transmitted packet will be successfully received by all such (connected) receivers. Time is slotted and only one packet can be transmitted at each time slot. Our system is static in the sense that no arrivals occur.

We now introduce the necessary notation. Integer $K$ ($K \leq F$) represents the coding window size, i.e., the number of packets that will be linearly combined using RLNC. $F$ is the file size and we assume that $K$ always exactly divides $F$ (i.e., $\frac{F}{K}$ is an integer). The $i^{th}$ batch refers to packets $(i-1)*K$ to $i*K-1$. Each batch contains $K$ packets which will be linearly combined/encoded to one packet. The number of batches is $b = F/K$, for a file of $F$ packets and coding window size $K$.

At each time slot, the base station selects a batch of $K$ packets to encode via RLNC. The encoded packet is then broadcast to the connected receivers. The batch to be encoded is based on a policy (Random Selection or Least Received). Such policies are described, in detail, later in this section.

Each receiver has a queue for storing the received encoded packets. As soon as a receiver receives $K$ packets, the packets are decoded and deleted from that queue. Linear independence of the encoded packets is assumed\footnote{Linear independence is justified due to a large enough field $\mathbb{F}_q$ from where the coefficients will be picked \cite{1}}. The coding overhead (the coefficients of the linear combinations must be transmitted with the packet) is considered negligible. As shown in \cite{1}, if the packet size $m$ is a lot greater than the coding window size $K$, this overhead can be ignored. Moreover, by using synchronized pseudo-random number generators at the sender and the receivers, this overhead can be made very small \cite{18}.

Furthermore, each receiver is assigned an attribute, namely the batch ID. This attribute represents the batch from which the receiver expects the encoded packets. At the beginning of the system (i.e. when $t=0$) the batch ID is set to 1, for all receivers. As soon as a receiver decodes a batch, its batch ID increases by 1. Any out of order packets (encoded packets of batch $i$ received by a receiver with batch ID $j$, where $i \neq j$) are discarded by the receiver.

When, at a given time slot, a subset of the connected receivers have successfully decoded a batch (received all $K$ encoded packets) that another disjoint subset of the connected receivers has yet to decode, a decision has to made by the base station as to which batch will be encoded and sent at that time slot. We propose and use the following two batch scheduling heuristic policies in order to dissolve these conflicts.
\begin{enumerate}
\item \textit{Random Selection (RS)} : This heuristic is based on randomly selecting one batch to encode. Each batch $i$ is selected with probability $\frac{N_i}{N_c}$, where $N_i$ is the number of connected receivers with batch ID $i$ and $N_c$ is the number of connected receivers. This heuristic will not be analysed - it is developed only for evaluation purposes in our experimentation process. 
\item \textit{Less Received (LR)} : This heuristic aims at balancing the queue size of the receivers. The batch ID of the receiver that has received the least number of packets defines the batch that will be selected for encoding (i.e., batch $i$ is selected when the receiver with the least received packets has batch ID $i$).
%\item Max Gain (MG) : This heuristic imitates the behaviour of a greedy algorithm. Each decision is locally optimal in the sense that it selects the batch that will yield the maximum gain, i.e. the one that would be useful to the greatest number of receivers. In case of a tie between batches, the one that would serve the receiver with the less received packets is selected.
\end{enumerate}
\begin{figure}

\centering
\includegraphics[scale = 0.45]{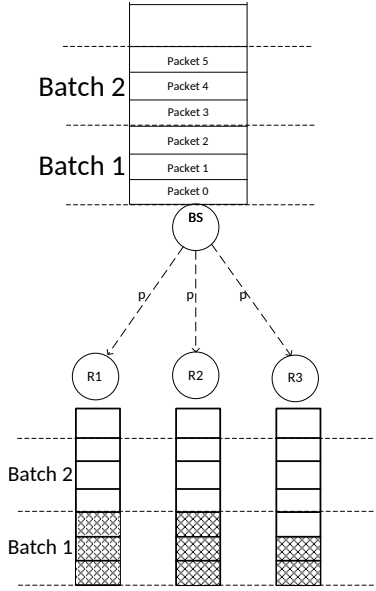}
\caption{System at time $t$. $K=3$, $N=3$}
\label{System}
\end{figure}
In figure \ref{System}, we show an example of our system at time slot $t$. The shaded boxes at the receivers represent the received packets. The coding window size, for this example, is 3. As we can see, receivers 1 and 2 have successfully decoded batch 1 (received 3 packets), thus their batch ID is 2. Receiver 3 has yet to decode batch 1 (received 2 packets), thus its batch ID is 1. Assume that in time slot $t+1$, all receivers are connected. When using the RS heuristic, batch 1 will be selected with probability $\frac{2}{3}$ and batch 2 will be selected with probability $\frac{1}{3}$. When using the LR heuristic, batch 1 will be selected, since R3 has the least received number of packets. Note here, that at time $t+2$, no conflicts will arise (when using the LR heuristic). %When MG is used, batch 2 will yield the maximum gain, thus this is the selected batch. If at time slot $t+1$ receivers 1 and 3 are connected, MG will select batch 1, since both batches have a gain of 1, but batch 1 will serve the receiver with the less received packets.

\section{Expected File Completion Time Computation}
When the coding window consists of the whole file ($K = F$), the expected completion time of the file can be found in \cite{1}. For our analysis we introduce the following :

Let $\mathbb{E}[T_F]$ be the mean completion time of a file of size $F$ under coding window $F$. $X_i$ is the additional slots needed for correct file reception by receiver $i$, due to possible channel disconnections ($X_i$ is also referred as failures), i.e. $X_i + F$ shows the total number of slots required by receiver $i$ to decode the file. The $X_i$'s are independent and identically distributed and follow a negative binomial distribution. Let $f_{X_i}(x)$ and $F_{X_i}(x)$ be the pdf and cdf of $X_i$, respectively and $Z = \max\limits_{1\leq i \leq N}X_i$. Then, \cite{1} has shown the following : 
\begin{center}
$
\mathbb{E}[T_F] = F + \mathbb{E}[Z] \stackrel{(a)}{=} F + \sum\limits_{z=0}^\infty P(Z > z) = 
$
\end{center}
\begin{center}
$
 F + \sum\limits_{z=0}^\infty (1 - P( \max\limits_{1\leq i \leq N}X_i \leq z)) \stackrel{(b)}{=} F + \sum\limits_{z=0}^\infty (1 - \prod\limits_{i=1}^N F_{X_i}(z)) 
$
\end{center}
\begin{center}
$ 
 = F + \sum\limits_{z=0}^\infty (1 - \prod\limits_{i=1}^N (\sum\limits_{t=0}^z f_{X_i}(t))) \stackrel{(c)}{\Rightarrow}
$
\end{center}
\begin{equation}
\label{NBIN}
\mathbb{E}[T_F] = F + \sum\limits_{z=0}^\infty (1 - (\sum\limits_{t=0}^z f_{X_1}(t))^N)
\end{equation}
The following notes are in order : 
(a) is true since $X_i$'s (and $Z$) are non-negative discrete random variables, 
(b) is true due to independence of $X_i$'s and
(c) is valid because the channel is symmetric, i.e. all the receivers have the same probability of being ON.

Eq. (\ref{NBIN}) refers to the completion time of the file if we encode over the whole file. By analogy, if $K=F$, $\mathbb{E}[T_K]$ will represent the completion time of one batch of $K$ packets. For our LR batch scheduling heuristic we define $\mathbb{E}[T_K^F]$ to be the completion time of the file under a coding window size $K$. We let $b = \frac{F}{K}$ and argue that

\begin{equation}
\label{bT}
\mathbb{E}[T_K^F] \approx b * \mathbb{E}[T_K]
\end{equation}

The justification is as follows : Formula \ref{bT} would be exact if the base station will transmit an encoded packet from batch $i$, \textit{only when all receivers have successfully decoded batch $i-1$} (as a counter example consider the case of Figure \ref{System} where at $t+1$ R3 is OFF and R1,R2 are ON). Furthermore, we note that the LR heuristic aims at balancing the receiver queues. In a perfectly balanced system and when $p=1$, the differences between the batch IDs will be zero (i.e. all receivers will receiver batch $i$ at the same time slot). However, when $p < 1$, the LR heuristic gives priority to the receiver with the least number of received packets, thus striving to balance the system through minimizing differences between the batch IDs of the receivers.

Furthermore, we should stress that once the file is segmented into blocks and RLNC is applied at each block, the random variables representing the file transfer completion time for each receiver $i$ are not independent. 
In order to proceed with our analysis, we will assume independence. The simulation results, in section V, demonstrate that our approximation represents the completion time of the file reasonably accurate.
%To justify \ref{bT}, we need a deeper understanding of both the formula and the heuristic. 
%\begin{itemize}
%\item $\mathbb{E}[T_F^F]$ : This formula assumes that the base station will transmit an encoded packet from batch $i$, only when all receivers have successfully decoded batch $i-1$.
%\item LR heuristic : This heuristic balances the queues. This means that priority will be given to the receiver that has received the least number of packets. After one (or more) receivers successfully receive batch $i$, they will receive an encoded packet from batch $i+1$ (before the rest of the receivers successfully decode batch $i$ ) only when all the connected receivers have received batch $i$.
%
%\end{itemize}
\subsection{Approximations for Expected File Completion Time}

Based on the Central Limit Theorem, a negative binomial random variable (representing the number of failures until the correct reception of $K$ packets), with parameters $K$ and $p$, can be approximated by a Gaussian random variable, for large $K$\footnote{The accuracy of the approximation increases as we increase $K$.} and moderate $p$\footnote{In our numerical simulations, section V, we assume $p \in [0.2, 0.8]$}\cite{20}. 

The approximating Gaussian random variable will have a mean value of $\frac{K}{p}-K$ and a standard deviation of $\sigma = \sqrt{\frac{K(1-p)}{p^2}}$ and it will represent the number of failures until the completion of the file transfer. Since we are interested in calculating the mean number of slots (trials) for the file transfer completion time, a constant $K$ can be added to the Gaussian random variable. Thus, the new, shifted, random variable will remain Gaussian with a mean value of $\mu = \frac{K}{p}$.

The approximating Gaussian variable $X$ represents the number of total slots for the transfer file completion time, for a single receiver (in a symmetric system). Therefore, \textit{$X$ is assumed to be positive} (with the negative tail carrying negligible probability). Then by following the same steps as for eq. (\ref{NBIN}) (for $F=K$), we get :
\begin{center}
$\mathbb{E}[T_K] = \int_{0}^\infty (1 - (\int_{0}^z f_{X}(t) dt)^N) dz \approx$
\end{center}
\begin{equation}
\label{Gauss}
 \int_{0}^\infty (1 - (F_X(z))^N) dz, 
\end{equation}
where $X$ is a Gaussian random variable with $\mu = \frac{K}{p}$, $\sigma = \sqrt{\frac{K(1-p)}{p^2}}$, $f_X(t)$ and $F_X(z)$ represent the pdf and cdf of the Gaussian random variable $X$, respectively.

In order to simplify eq. \ref{Gauss}, we assume the following :\\
For a Gaussian random variable $X$, $F_X(\mu + n\sigma) - F_X(\mu - n\sigma) = erf(\frac{n}{\sqrt{2}})$. When $erf(\frac{n}{\sqrt{2}}) \approx 1$, then $F_X(\mu + n\sigma) \approx 1$ and $F_X(\mu - n\sigma) \approx 0$. In our case, we wish to find $\widetilde{n}$ such that when $z \geq \mu + \widetilde{n}\sigma$ then $(F_X(z))^N \geq 1-\alpha$, for small $\alpha$; typically in our study $\alpha \approx 0.01$. Thus, the desired value $\widetilde{n}$ will need to satisfy : 
\begin{equation}
\label{nTilde}
\widetilde{n} = \min_{n}\{(erf(\frac{n}{\sqrt{2}}))^N \geq 0.99\}
\end{equation}
The range of $X$ will be within $\mu \pm \widetilde{n}\sigma$. Furthermore, since $X$ is assumed to be positive, $\widetilde{n}$ should satisfy $\mu - \widetilde{n}\sigma > 0$.

Based on the aforementioned assumptions, we can conclude the following :
\begin{enumerate}[label=(\alph*)]
\item For the desired value $\widetilde{n}$, $F_X(z) \approx 1$ (and also $(F_X(z))^N \approx 1$), for $z \geq \mu + \widetilde{n}\sigma$.
\item By symmetry, $F_X(z) \approx 0$ (and $(F_X(z))^N \approx 0$), for $z \leq \mu - \widetilde{n}\sigma$.
\end{enumerate}
Then from eq. \ref{Gauss} and \ref{nTilde} and using facts $(a)$ and $(b)$, we get : 
\begin{center}
$
\mathbb{E}[T_K] \stackrel{(a)}{\approx} \int_{0}^{\mu + \widetilde{n}\sigma} (1 - (F_X(z))^N) dz \stackrel{(b)}{=} 
$\\
\end{center}
\begin{center}
$\int_0^{\mu - \widetilde{n}\sigma} 1 dz + \int_{\mu - \widetilde{n}\sigma}^{\mu + \widetilde{n}\sigma} (1 - (F_X(z))^N) dz = $\\
\end{center}
\begin{center}
$ \mu - \widetilde{n}\sigma + \int_{\mu - \widetilde{n}\sigma}^{\mu + \widetilde{n}\sigma} (1 - (\int_{0}^z f_{X}(t) dt)^N) dz = $
\end{center}
\begin{center}
$ \mu - \widetilde{n}\sigma + \int_{\mu - \widetilde{n}\sigma}^{\mu + \widetilde{n}\sigma} (1 - (\int_0^{\mu - \widetilde{n}\sigma} f_{X}(t) dt + \int_{\mu - \widetilde{n}\sigma}^z f_{X}(t) dt)^N) dz$
\end{center}
\begin{center}
$\approx \mu - \widetilde{n}\sigma + \int_{\mu - \widetilde{n}\sigma}^{\mu + \widetilde{n}\sigma} (1 - (F_X(\mu - \widetilde{n}\sigma) + \int_{\mu - \widetilde{n}\sigma}^z f_{X}(t) dt)^N) dz$
\end{center}
\begin{center}
$\stackrel{(b)}{\approx} \mu - \widetilde{n}\sigma + \int_{\mu - \widetilde{n}\sigma}^{\mu + \widetilde{n}\sigma} (1 - \int_{\mu - \widetilde{n}\sigma}^z f_{X}(t) dt)^N) dz$
\end{center}
Therefore,
\begin{equation}
\label{GaussAppr}
\mathbb{E}[T_K] \approx \mu + \widetilde{n}\sigma - \int_{\mu - \widetilde{n}\sigma}^{\mu + \widetilde{n}\sigma} ((\int_{\mu - \widetilde{n}\sigma}^z f_{X}(t) dt)^N) dz,
\end{equation}
when assuming $\mu - \widetilde{n}\sigma > 0$. This condition will be true for a given $\widetilde{n}$ from eq. \ref{nTilde} and a particular selection of $K$ as follows: 
\begin{equation}
\label{Kmin}
\frac{K}{p} - \widetilde{n}\sqrt{\frac{K(1-p)}{p^2}} > 0 \Rightarrow K > \widetilde{n}^2(1-p)
\end{equation}

\begin{table}[H]
\centering
\adjustbox{max width=\columnwidth}{
\begin{tabular}{ c | c | c |}
\cline{2-3}
  \multicolumn{1}{r||}{}  	& $N \leq 3$, & $4 \leq N \leq 158$\\ \cline{2-3}
   
  \multicolumn{1}{r||}{}	& $\widetilde{n}=3$ & $\widetilde{n}=4$ \\ \hline \hline
  \multicolumn{1}{|r||}{$p = 0.2$}  & $K=8$  	& $K=13$  \\ \hline
  \multicolumn{1}{|r||}{$p = 0.4$}  & $K=6$ 	& $K=10$  \\ \hline
  \multicolumn{1}{|r||}{$p = 0.6$}  & $K=4$ 	& $K=7$   \\ \hline
  \multicolumn{1}{|r||}{$p = 0.7$}  & $K=3$	& $K=5$   \\ \hline
  \multicolumn{1}{|r||}{$p = 0.8$}  & $K=2$ 	& $K=4$   \\ \hline
\end{tabular}}
\caption{Minimum Coding Window Size for satisfying the constraint (\ref{Kmin}).}
\label{m-ns}
\end{table}

In table \ref{m-ns}, in the second row, we can see the value of $\widetilde{n}$ for each $N$. For those values, and for each $p$, we can see the minimum coding window that satisfies the constraint (\ref{Kmin}). 
We observe that the smallest coding window size required to satisfy the constraint (\ref{Kmin}) is small enough for all values of $\widetilde{n}$ and $p$. In our subsequent analysis, we assume that $K$ is equal or larger than the values shown in table \ref{m-ns}.
\section{Delay Constraints and Selection of Coding Window Size ($K$)}

When $K$ satisfies the constraint (\ref{Kmin}), the expected file transfer completion time of the LR heuristic is given by eq. (\ref{bT}). In this case, our goal is to find the smallest Coding Window size $K$ that results into an acceptable delay. Coding over the whole file ($K=F$) will minimize the file transfer completion time (\cite{18}). We denote this by $\mathbb{E}[T_{opt}] \stackrel{\vartriangle}{=} \mathbb{E}[T_F]$, and can be found by eq. (\ref{NBIN}). From our experimentation (in section V), we observed that near optimal file transfer completion time can be achieved using an appropriate coding window size $K \ll F$. Therefore, given $\epsilon > 0$, we wish to determine $K$ so that : 
\begin{equation}
\label{epsilon}
\frac{\mathbb{E}[T_K^F] - \mathbb{E}[T_{opt}]}{\mathbb{E}[T_{opt}]} \leq \epsilon,
\end{equation}
where $\epsilon$ is the user defined delay constraint in terms of a percentage of $\mathbb{E}[T_{opt}]$. As an example, we will see later (section V) that for $N=6$, $F=10000$ and $p=0.2$, for $\epsilon = 10\%$ the minimum coding window size is $K = 400$ ($4\%$ of $F$).

Starting from eq. (\ref{bT}) and using eq. (\ref{GaussAppr}), the completion time, under a coding window $K$, can be approximated as follows :
\begin{center}
$\mathbb{E}[T_K^F] = \mathbb{E}[T_K] *b = b(\mu_K + \widetilde{n}\sigma_K) - $
\end{center}
\begin{equation}
\label{T_K^F}
- b\int_{\mu_K - \widetilde{n}\sigma_K}^{\mu_K + \widetilde{n}\sigma_K} ((\int_{\mu_K - \widetilde{n}\sigma_K}^z \frac{1}{\sigma_K\sqrt{2\pi}}e^{-\frac{(t- \mu_K)^2}{2\sigma_K^2}} dt)^N) dz),
\end{equation}
where $\mu_K = \frac{K}{p}$, $\sigma_K = \sqrt{\frac{K(1-p)}{p^2}}$ and $b_K = \frac{F}{K}$.\\
Furthermore, the minimum completion time $\mathbb{E}[T_{opt}]$ can be also approximated by : 
\begin{center}
$\mathbb{E}[T_{opt}] = \mu_F + \widetilde{n}\sigma_F - $
\end{center}
\begin{equation}
\label{Topt}
- \int_{\mu_F - \widetilde{n}\sigma_F}^{\mu_F + \widetilde{n}\sigma_F} ((\int_{\mu_F - \widetilde{n}\sigma_F}^z \frac{1}{\sigma_F\sqrt{2\pi}}e^{-\frac{(t- \mu_F)^2}{2\sigma_F^2}} dt)^N) dz,
\end{equation}
where $\mu_F = \frac{F}{p}$ and $\sigma_F = \sqrt{\frac{F(1-p)}{p^2}}$. Since $F = Kb_K, \mu_F = b_K\mu_K$ and $ \sigma_F = \sqrt{b_K}\sigma_K $, we can substitute these values in eq. (\ref{T_K^F}) and (\ref{Topt}) and get\footnote{we suppress subscripts to simplify the presentation} :
\begin{equation}
\label{TFK}
\mathbb{E}[T_K^F] = b\mu + b\widetilde{n}\sigma - b\int_{\mu - \widetilde{n}\sigma}^{\mu + \widetilde{n}\sigma} ((\int_{\mu - \widetilde{n}\sigma}^z \frac{1}{\sigma\sqrt{2\pi}}e^{-\frac{(t- \mu)^2}{2\sigma^2}} dt)^N) dz)
\end{equation}
\begin{center}
$\mathbb{E}[T_{opt}] = b\mu  +  \sqrt{b}\widetilde{n}\sigma -$
\end{center}
\begin{equation}
\label{T}
- \int_{b\mu - \sqrt{b}\widetilde{n}\sigma}^{b\mu + \sqrt{b}\widetilde{n}\sigma} ((\int_{b\mu - \sqrt{b}\widetilde{n}\sigma}^z \frac{1}{\sigma\sqrt{2b\pi}}e^{-\frac{(t- b\mu)^2}{2b\sigma^2}} dt)^N) dz,
\end{equation}
where $\mu = \frac{K}{p}$, $\sigma = \sqrt{\frac{K(1-p)}{p^2}}$ and $b = \frac{F}{K}$.\\
Furthermore, it can be readily shown that : 
\begin{center}
$\int_{b\mu - \sqrt{b}\widetilde{n}\sigma}^{b\mu + \sqrt{b}\widetilde{n}\sigma} ((\int_{b\mu - \sqrt{b}\widetilde{n}\sigma}^z \frac{1}{\sigma\sqrt{2b\pi}}e^{-\frac{(t- b\mu)^2}{2b\sigma^2}} dt)^N) dz =  $\\
$\sqrt{b}\int_{\mu - \widetilde{n}\sigma}^{\mu + \widetilde{n}\sigma} ((\int_{\mu - \widetilde{n}\sigma}^z \frac{1}{\sigma\sqrt{2\pi}}e^{-\frac{(t- \mu)^2}{2\sigma^2}} dt)^N) dz$ and \\
$\int_{\mu - \widetilde{n}\sigma}^{\mu + \widetilde{n}\sigma} ((\int_{\mu - \widetilde{n}\sigma}^z \frac{1}{\sigma\sqrt{2\pi}}e^{-\frac{(t- \mu)^2}{2\sigma^2}} dt)^N) dz = $\\
$\sigma\int_{- \widetilde{n}}^{\widetilde{n}} ((\int_{- \widetilde{n}}^z \frac{1}{\sqrt{2\pi}}e^{-\frac{t^2}{2}} dt)^N) dz$
\end{center}

%
%Proof of \ref{eq7} : \\
%$\int_{b\mu - \sqrt{b}\widetilde{n}\sigma}^{b\mu + \sqrt{b}\widetilde{n}\sigma} ((\int_{b\mu - \sqrt{b}\widetilde{n}\sigma}^z \frac{1}{\sigma\sqrt{2b\pi}}e^{-\frac{(t- b\mu)^2}{2b\sigma^2}} dt)^N) dz = \int_{b\mu - \sqrt{b}\widetilde{n}\sigma}^{b\mu + \sqrt{b}\widetilde{n}\sigma} ((\int_{-\widetilde{n}\sigma}^{\frac{z}{\sqrt{b}} - \sqrt{b}\mu} -  \frac{1}{\sigma\sqrt{2b\pi}}e^{-\frac{k^2}{2\sigma^2}} \sqrt{b}dk)^N) dz $,\\ where $k = \frac{t - b\mu}{\sqrt{b}}$ \\
%$\int_{b\mu - \sqrt{b}\widetilde{n}\sigma}^{b\mu + \sqrt{b}\widetilde{n}\sigma} ((\int_{-\widetilde{n}\sigma}^{\frac{z}{\sqrt{b}} - \sqrt{b}\mu} -  \frac{1}{\sigma\sqrt{2\pi}}e^{-\frac{k^2}{2\sigma^2}} dk)^N) dz = \sqrt{b}\int_{\mu - \widetilde{n}\sigma}^{\mu + \widetilde{n}\sigma} ((\int_{-\widetilde{n}\sigma}^{w - \mu} -  \frac{1}{\sigma\sqrt{2\pi}}e^{-\frac{k^2}{2\sigma^2}} dk)^N) dw$, \\ where $w = \mu + \frac{z}{\sqrt{b}} - \sqrt{b}\mu$. Furthermore,\\
%$\int_{\mu - \widetilde{n}\sigma}^{\mu + \widetilde{n}\sigma} ((\int_{\mu - \widetilde{n}\sigma}^z \frac{1}{\sigma\sqrt{2\pi}}e^{-\frac{(t- \mu)^2}{2\sigma^2}} dt)^N) dz) = \int_{\mu - \widetilde{n}\sigma}^{\mu + \widetilde{n}\sigma} ((\int_{- \widetilde{n}\sigma}^{z-\mu} \frac{1}{\sigma\sqrt{2\pi}}e^{-\frac{k^2}{2\sigma^2}} dk)^N) dz)$, where $k = t -\mu$. \\
Let $A = \int_{- \widetilde{n}}^{\widetilde{n}} ((\int_{- \widetilde{n}}^z \frac{1}{\sqrt{2\pi}}e^{-\frac{t^2}{2}} dt)^N) dz$. Eq. (\ref{TFK}) and (\ref{T}) can be rewritten as :
\begin{equation}
\label{9}
\mathbb{E}[T_K^F] =  b\mu + b\widetilde{n}\sigma - b\sigma A
\end{equation}
\begin{equation}
\label{10}
\mathbb{E}[T_{opt}] =  b\mu + \sqrt{b}\widetilde{n}\sigma - \sqrt{b}\sigma A
\end{equation}

Substituting \ref{9} and \ref{10} into \ref{epsilon} and noting that $b$ is a function of $K$ and $A$, $\widetilde{n}$ are functions of $N$, we have the following lemma for the selection of the coding window size $K$. 

\textsc{Lemma} : For given $F$, $N$, $p$ and $\epsilon > 0$, the coding window size $K$ that results in eq. \ref{epsilon} satisfies : 
\begin{equation}
\label{epsilon1}
\frac{\sqrt{1 - p}(\widetilde{n} - A)}{\sqrt{F} + \sqrt{1 - p}(\widetilde{n} - A)}(\sqrt{b_K} - 1) \leq \epsilon, 
\end{equation}
\begin{center}
where $b_K=\frac{F}{K}$,  $\widetilde{n} = \min_{n}\{(erf(\frac{n}{\sqrt{2}}))^N \geq 0.99\},$\\$A = \int_{- \widetilde{n}}^{\widetilde{n}} ((\int_{- \widetilde{n}}^z \frac{1}{\sqrt{2\pi}}e^{-\frac{t^2}{2}} dt)^N) dz$
\end{center}
Therefore, the smallest $K$ that satisfies eq. (\ref{epsilon1}) will also minimize the decoding delay at the receivers end while achieving \textit{near optimal} file transfer completion time (i.e., according to (\ref{epsilon})). Thus, \textit{this value of $K$ will result in balancing the file transfer completion time and the decoding complexity/delay at the receivers end}. Moreover, eq. (\ref{epsilon1}) shows that, \textit{ under suitable conditions (large enough $F$) the minimum required coding window size is the same regardless of the file size}. Furthermore, by using a much smaller coding window size $K$ as compared to the entire file size $F$, packets of the earlier blocks are made available to receivers (for delivery to higher protocol layers or forwarding) much faster.

\section{Experiments - Results}

We performed our simulations with various values for all the system parameters to verify the correctness of the above formulas. Below is an overview of all those values :
\begin{itemize}
\item Number of Receivers ($N$) : 3, 6, 12, 25, 50, 100
\item File Size in Packets ($F$) : 400, 500, 1000, 1500, 2000, 2500, 5000, 10000
\item Coding Window Size ($K$) : for each file size, the coding window takes as values all the integer numbers that fully divide the file size, beginning from 2. 
\item Connectivity probability $p$ is assumed to be the same for all receivers and is equal to : 0.2, 0.4, 0.6, 0.7, 0.8.
\end{itemize}

\begin{figure}[H]

\includegraphics[scale=.40]{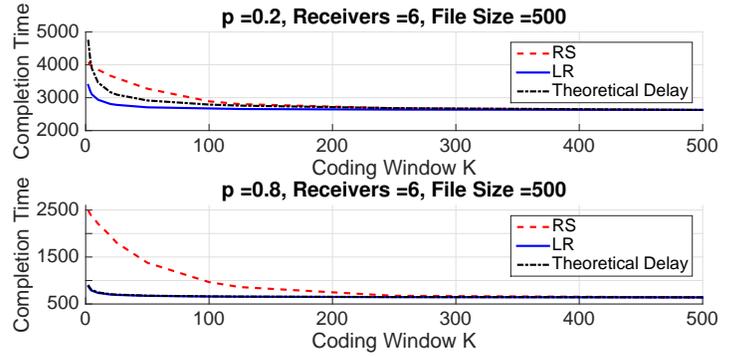}
\caption{Heuristics vs equation \ref{bT} when $p$ increases}
\label{inc_p}
\end{figure}

\begin{figure}[H]
\includegraphics[scale=.40]{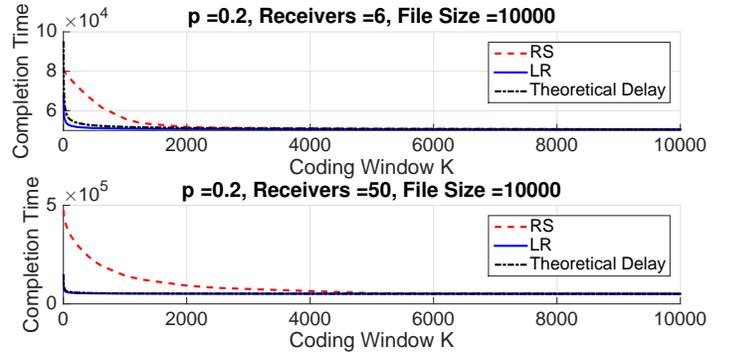}
\caption{Heuristics vs equation \ref{bT} when $N$ increases}
\label{inc_r}
\end{figure}

\begin{figure}[H]
\includegraphics[scale=.40]{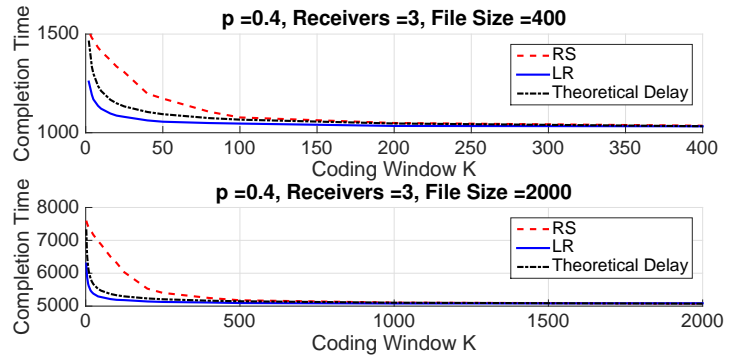}
\caption{Heuristics vs equation \ref{bT} when $F$ increases}
\label{inc_f}
\end{figure}

Due to space restriction, we now present some representative results. Figures \ref{inc_p} - \ref{inc_f} show the completion time of the file versus the coding window size $K$. In each figure two out of the three parameters ($N$, $F$ and $p$) were kept constant and we compared the improvement in the accuracy of eq. (\ref{bT}) (Theoretical Delay), when the third parameter was varied. It can be seen that eq. (\ref{bT}) represents the completion time of the LR heuristic, reasonably accurate. % In those figures we also calculated the minimum coding window size in order to achieve completion time $\epsilon$ times larger than $\mathbb{E}[T_{opt}]$, for $\epsilon = 10\%, 5\%, 1\%, 0.5\%$. These values can been seen in the figures (red coloured points).
In general, we observe that $\mathbb{E}[T_K^F]$ rapidly decreases as a function of $K$ and closely approaches $\mathbb{E}[T_{opt}]$ for values of coding window size $K$ that are a fraction of $F$. As we mentioned in the introduction, this is desirable because, for smaller coding window size $K$, \textit{the decoding delay is significantly less with a cost of slight increase of the file transfer completion time}. Additionally, \textit{the storage and computational requirements for the receivers are relaxed}, since each receiver only needs to store a maximum of $K$ packets and solve a $K \times K$ linear system (by Gaussian elimination). Furthermore, it is worth noting that the LR heuristic largely outperforms the Random Selection (RS) heuristic.
% and also in the tables \ref{0.2-0.8} - \ref{400-2K}.

\begin{figure}[H]
\centering
\includegraphics[scale=.40]{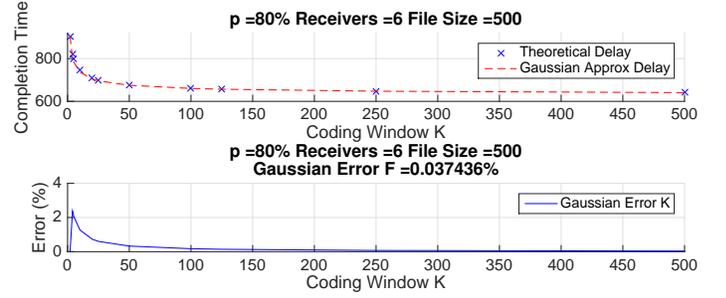}
\caption{Gaussian Approximation vs Negative Binomial and approximation errors - high $p$}
\label{inc_p1}
\end{figure}

\begin{figure}[H]
\centering
\includegraphics[scale=.40]{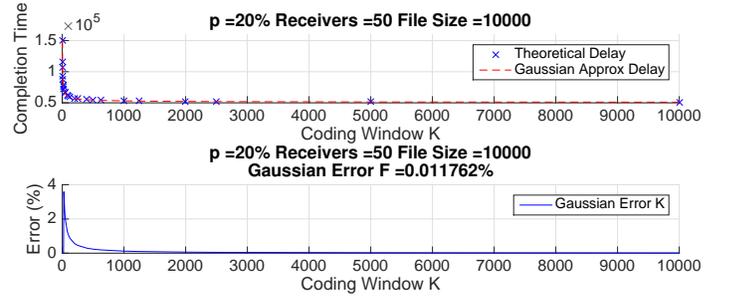}
\caption{Gaussian Approximation vs Negative Binomial and approximation errors - low $p$}
\label{inc_r1}
\end{figure}

\begin{figure}[H]
\includegraphics[scale=.4]{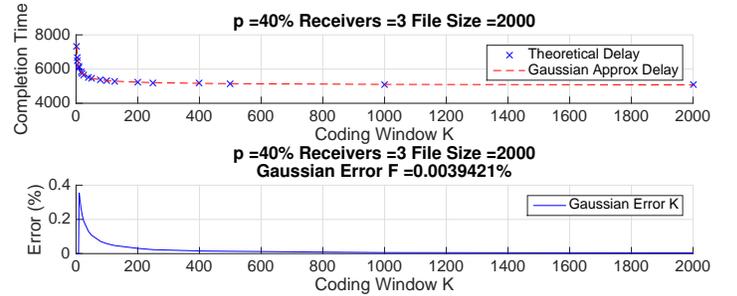}
\caption{Gaussian Approximation vs Negative Binomial and approximation errors - mid $p$}
\label{inc_f1}
\end{figure}

In the upper part of figures \ref{inc_p1} - \ref{inc_f1}, we plot the completion time based on the Gaussian Approximation (eq. (\ref{9})) vs  the completion time based on the negative binomial random variable (eq. (\ref{bT})). As we can see, the Gaussian approximation represents the negative binomial reasonably accurate, especially when $K$ is increased. In the lower graphs within the same figures we can see the approximation errors.  The solid lines represent the percent difference between eq. (\ref{9}) and (\ref{bT}) and the Gaussian Error $F$ represents the percent difference between eq. (\ref{10}) and (\ref{NBIN}). The error of the approximation of $\mathbb{E}[T_{opt}]$ is reasonably low with a maximum value of 0.36\% in our experiments. The corresponding maximum error of the approximation of $\mathbb{E}[T_K^F]$ increases somehow (up to 8\% in some experiments). But as we can see, the error rapidly drops in levels much below 1\% as $K$ grows a little larger. %In all the cases, this error drops below 1\% approximately when the completion time rapidly decreases. 
The approximations were derived in order to find a balance between the completion time and decoding delay under RLNC. The values of the coding window sizes for which the error percentage is high (above 1\%) will not affect our objective, because as we can see from figures \ref{inc_p1} - \ref{inc_f1}, for such values $K$, the file transfer completion time is far from $\mathbb{E}[T_{opt}]$.

Tables \ref{0.2-0.8} - \ref{400-2K} show the percentage of $F$ (the coding window $K$ is expressed as a percentage of the file size) that can be used as a coding window size in order to achieve completion time of at most $(1+\epsilon)\mathbb{E}[T_{opt}]$. Each cell of the tables contains two entries, one for each set of experiments shown in figures \ref{inc_p} - \ref{inc_f}. The first two columns are the $F \%$ for our heuristics. The third, fourth and fifth column are derived using the formulas \ref{bT}, \ref{9} and \ref{epsilon1} respectively. As we can see from table \ref{0.2-0.8}, when increasing $p$, the $F \%$ of the heuristics increases whereas those of the theoretical formulas decrease. Tables \ref{6-50} and \ref{400-2K} show a similar trend - when increasing the number of receivers, the $F \%$ is increasing and when increasing the file size, the percentage is decreasing. The latter is \textit{the essential contribution of our work, showing that for large files, coding window sizes that are a small fraction of $F$ can achieve near optimal file transfer completion time}. 

Moreover, from tables \ref{0.2-0.8} to \ref{400-2K}, we can see that as either $p$ or $N$ or $F$ increases the difference between the theoretical results (last three columns) and our heuristic LR decreases. This is of great importance, since it shows that as the load of the system increases, the accuracy of our approximation increases. Furthermore, it is evident, from table \ref{6-50}, that coding is preferable in large systems. An increase of almost 9 times at the number of the receivers results in an increase of a maximum of 3 times in the minimum coding window size. Table \ref{largeF}, shows the minimum coding window size (in packets and as a percentage of $F$) to achieve completion time of at most $(1+\epsilon)\mathbb{E}[T_{opt}]$ for large file sizes. As we can see, in most cases, the minimum required coding window size is the same regardless of the file size. In the rest of the cases, the differences mainly occur because not all file sizes have the same possible coding windows (i.e. $F/K$ must be an integer).

\begin{table}[H]
\large
\centering
\adjustbox{max width=\columnwidth}{
\begin{tabular}{| c || cc | cc | cc | cc | cc |}
\hline
   $\epsilon$ & \multicolumn{2}{c|}{RS} & \multicolumn{2}{c|}{LR} & \multicolumn{2}{c|}{Eq (\ref{bT})} & \multicolumn{2}{c|}{Eq. (\ref{9})} & \multicolumn{2}{c|}{Eq. (\ref{epsilon1})}\\ 
   \hline \hline
   & \multicolumn{2}{c|}{$p$}   & \multicolumn{2}{c|}{$p$}   & \multicolumn{2}{c|}{$p$}   & \multicolumn{2}{c|}{$p$}   & \multicolumn{2}{c|}{$p$}\\
   & $0.2$ & $0.8$& $0.2$ & $0.8$& $0.2$ & $0.8$& $0.2$ & $0.8$& $0.2$ & $0.8$ \\ \hline
  $10\%$  & 25\% & 50\%   & 4\% & 4\%  & 20\% & 5\%   &20\% & 4\%   &20\% & 4\% \\ \hline
  $5\%$   & 50\% & 100\%   & 10\% & 20\%   & 25\% & 20\%     &25\% & 20\%     &25\% & 20\% \\ \hline
  $1\%$   & 100\% & 100\%  & 50\% & 50\%     & 100\% & 100\%    &100\% & 100\%    &100\% & 100\% \\ \hline
  $0.5\%$ & 100\% & 100\%  & 50\% & 100\%    & 100\% & 100\%    &100\%  & 100\%   &100\% & 100\%\\ \hline
\end{tabular}}
\caption{Percentage of minimum Coding Window Size - $p = 0.2$ \& $0.8, N = 6, F = 500$}
\label{0.2-0.8}
\end{table}

\begin{table}[H]
\large
\centering
\adjustbox{max width=\columnwidth}{
\begin{tabular}{| c || cc | cc | cc | cc | cc | }
\hline
 $\epsilon$ & \multicolumn{2}{c|}{RS} & \multicolumn{2}{c|}{LR} & \multicolumn{2}{c|}{Eq (\ref{bT})} & \multicolumn{2}{c|}{Eq. (\ref{9})} & \multicolumn{2}{c|}{Eq. (\ref{epsilon1})}\\ 
   \hline \hline
   & \multicolumn{2}{c|}{$N$}& \multicolumn{2}{c|}{$N$}& \multicolumn{2}{c|}{$N$}& \multicolumn{2}{c|}{$N$}& \multicolumn{2}{c|}{$N$}\\
   & $6$ & $50$ & $6$ & $50$& $6$ & $50$& $6$ & $50$& $6$ & $50$ \\ \hline
   $10\%$  & 12.5\% & 50\%   & 0.25\% & 2\%  & 1.25\% & 4\%   &1.25\% & 4\%   &1.25\% & 4\% \\ \hline
  $5\%$   & 20\% & 100\%   & 0.8\% & 5\%   & 4\% & 10\%      &4\% & 10\%     &4\% & 10\% \\ \hline
  $1\%$   & 50\% & 100\%  & 10\% & 50\%     & 50\% & 50\%    &50\% & 50\%    &50\% & 50\% \\ \hline
  $0.5\%$ & 100\% & 100\%  & 20\% & 100\%    & 50\% & 100\%    &50\% & 100\%    &50\% & 100\% \\ \hline
\end{tabular}}
\caption{Percentage of minimum Coding Window Size - $p = 0.2, N = 6$ \& $50, F = 10000$}
\label{6-50}
\end{table}

\begin{table}[H]
\large
\centering
\adjustbox{max width=\columnwidth}{
\begin{tabular}{| c || cc | cc | cc | cc | cc | }
\hline
    $\epsilon$ & \multicolumn{2}{c|}{ RS} & \multicolumn{2}{c|}{ LR} & \multicolumn{2}{c|}{Eq (\ref{bT})} & \multicolumn{2}{c|}{Eq. (\ref{9})} & \multicolumn{2}{c|}{Eq. (\ref{epsilon1})}\\ 
   \hline \hline
   & \multicolumn{2}{c|}{ $F$}& \multicolumn{2}{c|}{ $F$}& \multicolumn{2}{c|}{ $F$}& \multicolumn{2}{c|}{ $F$}& \multicolumn{2}{c|}{ $F$} \\ 
   & $0.4K$ & $2K$ & $0.4K$ & $2K$& $0.4K$ & $2K$& $0.4K$ & $2K$& $0.4K$ & $2K$ \\ \hline
   $10\%$  &  20\% &  10\%   &  2.5\% &  0.8\%  &  6.25\% &  2\%   & 6.25\% &  2\%   & 6.25\% &  2\% \\ \hline
   $5\%$  &  25\% &  20\%   &  6.25\% &  2\%   &  20\% &  6.25\%     & 20\% &  6.25\%     & 20\% &  6.25\% \\ \hline
   $1\%$ &  100\% &  50\%  &  50\% &  12.5\%     &  100\% &  50\%    & 100\% &  50\%    & 100\% &  50\% \\ \hline
   $0.5\%$ &  100\% &  100\%  &  50\% &  20\%    &  100\% &  100\%    & 100\% &  100\%    & 100\% &  100\% \\ \hline
\end{tabular}}
\caption{Percentage of minimum Coding Window Size - $p = 0.4, N = 3, F = 400$ \& $2000$}
\label{400-2K}
\end{table}

\begin{table}[H]
\fontsize{8}{6}\selectfont
\centering
\adjustbox{max width=\columnwidth}{
\begin{tabular}{| c || c | c | c | c | }
\hline
    $\epsilon$ & $F = 8K$ & $F = 10K$ & $F = 12K$ & $F = 14K$ \\ 
   \hline \hline
   $10\%$  &  160 (2\%) &  200 (2\%)   &  160 (1.33\%) &  175 (1.25\%)\\ \hline
   $5\%$  &  500 (6.25\%) &  500 (5\%)   &  500 (4.17\%) &  560 (4\%)   \\ \hline
   $1\%$ &  4000 (50\%) &  5000 (50\%)  &  4000 (33.3\%) &  7000 (50\%)  \\ \hline
   $0.5\%$ &  8000 (100\%) &  10000 (100\%)  &  12000 (100\%) &  7000 (50\%) \\ \hline
\end{tabular}}
\caption{Minimum Coding Window Size (and percentage) - $p = 0.6, N = 50$}
\label{largeF}
\end{table}

\section{Conclusions}
In this paper, we applied Random Linear Network Coding in a single-hop network for broadcast communications where a base station transmits one file to $N$ receivers. We presented two batch scheduling heuristic policies namely Random Selection (RS) and Least Received (LR). The performance of the LR was approximated and simulated. Furthermore, we provided a formula for balancing the file transfer completion time and decoding delay based on a user defined delay constraint, for the LR heuristic. We concluded that moderate coding window size $K$ can achieve almost optimal performance using the LR heuristic policy. By using a moderate/smaller coding window size, the decoding complexity and delay at the receivers ends can decrease substantially. Moreover, we showed that for large enough files, the coding window size that achieves near optimal performance is the same regardless of the file size.

Our future research will be focused on finding the characteristics of the optimal scheduling policy. Furthermore, the asymptotic performance of the LR as $N$, $F$, $p$ grows large will be investigated.

\bibliography{referencesCamReady}
\bibliographystyle{IEEETran}

\end{document}